\documentclass[twocolumn]{aastex631}
\usepackage[T1]{fontenc}
\usepackage{amsmath}

\begin{document}

\title{Unified Solar Modulation Potential for Same-Charge Cosmic Rays and Implications for Local Interstellar Spectra}

\author{Meng-Jie Zhao}
\email{zhaomj@ihep.ac.cn}
 \affiliation{%
 Key Laboratory of Particle Astrophysics, Institute of High Energy Physics, Chinese Academy of Sciences, Beijing 100049, China}
\affiliation{
China Center of Advanced Science and Technology, Beijing 100190, China 
}%
 \author{Xing-Jian Lv}
\email{lvxj@ihep.ac.cn}
 \affiliation{%
 Key Laboratory of Particle Astrophysics, Institute of High Energy Physics, Chinese Academy of Sciences, Beijing 100049, China}
\affiliation{
 University of Chinese Academy of Sciences, Beijing 100049, China 
}%
 \author{Xiao-Jun Bi}
 \email{bixj@ihep.ac.cn}
\affiliation{%
 Key Laboratory of Particle Astrophysics, Institute of High Energy Physics, Chinese Academy of Sciences, Beijing 100049, China}
\affiliation{
 University of Chinese Academy of Sciences, Beijing 100049, China 
}%
\author{Kun Fang}
\email{fangkun@ihep.ac.cn}
\affiliation{%
 Key Laboratory of Particle Astrophysics, Institute of High Energy Physics, Chinese Academy of Sciences, Beijing 100049, China}
 \author{Peng-Fei Yin}
\email{yinpf@ihep.ac.cn}
\affiliation{%
 Key Laboratory of Particle Astrophysics, Institute of High Energy Physics, Chinese Academy of Sciences, Beijing 100049, China}



\begin{abstract}
The energy spectra of cosmic rays (CRs) below tens of GeV are significantly modulated by solar activity within the heliosphere. To investigate the properties of Galactic CRs, it is important to determine the unmodulated local interstellar spectrum (LIS). Recent high-precision temporal measurements of CR energy spectra, released by the AMS-02 collaboration, provide a crucial observational foundation for this endeavor. In this study, we employ the widely used force-field approximation (FFA) model to analyze the AMS-02 data, and attempt to derive the LIS for CR protons and positrons. By applying a non-LIS method, we derive temporal variations of the relative solar modulation potential, $\Delta\phi$, for individual particle species. Our analysis demonstrates that the FFA provides sufficient accuracy in explaining the AMS-02 spectral measurements of all particles during the low solar activity period. Notably, the derived $\Delta\phi(t)$ for protons and positrons, as well as for electrons and antiprotons, exhibit excellent consistency, indicating that particles with the same charge sign can be effectively described within a unified FFA framework during the low solar activity period. Having established a well-constrained proton LIS and its associated modulation potential, we apply the common modulation behavior between positrons and protons to demodulate the AMS-02 positron measurements, and derive the positron LIS without necessitating prior knowledge of its characteristics. This LIS is useful for quantitative investigations into potential exotic origins of CR positrons. 
\end{abstract}



\section{Introduction} \label{sec:intro}

The interaction between Cosmic rays (CRs) and the solar environment, referred to as solar modulation, plays an important role in shaping the observed properties of these particles.
Within the heliosphere, the turbulent solar wind and embedded magnetic field significantly alter the energy and trajectory of CRs at low energies. Consequently, the CR spectra observed at the top of Earth's atmosphere exhibit deviations from those in local interstellar space, namely the local interstellar spectra (LIS). The characteristics of the observed CR spectra arise from the intricate interplay of three key processes: the production of CRs, their propagation through the Galactic medium, and the solar modulation within the heliosphere. Therefore, to investigate the properties of Galactic CRs, it is essential to reasonably describe the solar modulation, enabling the extraction of the CR LIS. Thanks to the high-quality measurements by Voyager~\citep{Cummings:2016pdr, 2019NatAs...3.1013S}, we now have a relatively good understanding of the proton LIS~\citep{Potgieter:2017zat, Fiandrini:2020puf}. However, there remains considerable uncertainties regarding the LIS of other CR particles (e.g., antiprotons and positrons).

The CR and particle physics communities have widely employed the force-field approximation~\citep{Gleeson:1968zza} (FFA), a simplified yet effective model, to characterize the solar modulation. The FFA is described by a single parameter, the solar modulation potential $\phi$, which is expected to vary temporally and exhibit a strong correlation with solar activity. 
Considering the distinct drift behaviors of particles with opposite charge signs, it is reasonable to assume different values of $\phi$ for particles based on their charge signs~\citep{Cholis:2015gna, Cholis:2020tpi, Kuhlen:2019hqb}.
However, in previous studies~\citep{Lipari:2018usj, Silver:2024ero} aimed at explaining observations through Galactic CR propagation calculations, FFA potentials for CRs with the same charge were often treated independently.
Despite the intricacies of realistic solar modulation processes, the FFA has demonstrated its efficacy in explaining the temporal flux variations of CR protons during periods of low solar activity (LSA)~\citep{Koldobskiy:2019olx, Wang:2019xtu,Corti:2019jlt,Zhu:2024ega}. This opens up a window for inferring the CR LIS without resorting to complex calculations.


Recently, the AMS-02 collaboration has released precise temporal flux measurements for two pairs of elementary particles: protons ($p$) and anti-protons ($\bar{p}$), as well as electrons ($e^-$) and positrons ($e^+$), spanning an 11-year solar cycle from May 2011 to June 2022~\citep{AMS:2025npj}. 
Using these valuable data, we can examine three critical questions: 1) Is the FFA still applicable when dealing with long-period, high-precision spectral measurements? 2) Can CR particles with the same charge sign be accurately characterized by a single FFA potential? 3) Is it feasible to extract the LIS with precision from the current observational results?
If affirmative answers to these questions can be established, it would undoubtedly impose stronger constraints on models of CR production and propagation.

This study is organized as follows. In Sec.~\ref{sec:FFA}, we introduce the framework of the FFA model and demonstrate its continued validity in explaining the time-varying measurements. In Sec.~\ref{sec:Non-LIS}, we introduce the Non-LIS method, which enables us to interpret the temporal structures without being affected by LIS uncertainties. In Sec.~\ref{sec:result1}, we fit the temporal data of AMS-02 for four types of CRs, demonstrating the validity of FFA during the low solar activity period and the commonness of $\phi(t)$ for particles with identical charge. In Sec.~\ref{sec:result2}, we first obtain the LIS and the total $\phi(t)$ of protons, and then demodulate the $e^+$ measurements to derive its LIS with the total $\phi(t)$. Finally, Sec.~\ref{sec:Conclusion} summarizes our findings.


\section{Force-Field modulation} \label{sec:FFA}
The FFA model is derived from the Parker transport equation~\citep{Parker:1965ejd} under assumptions such as the steady state condition, spherical symmetry, and neglect of adiabatic expansion~\citep{Gleeson:1968zza}. The discussion of its limitations can be found in \citet{2004JGRA..109.1101C}.
After the FFA modulation, the kinetic energy per nucleon ($E_{k/n}$ or $E$ for short) of each charged particle is reduced by a quantity $\frac{|Z|e}{A}\phi$, where $\phi$ is the modulation potential (typically assumed to be energy-independent), Z and A are the charge and mass numbers of CRs, and $|Z|e$ is the absolute charge. The modulated CR spectrum can be expressed as:
\begin{equation}\label{eq:FFA}
   \Phi\left(E\right)=\frac{E(E+2m_0)}{(E+\frac{|Z|e}{A}\phi)(E+\frac{|Z|e}{A}\phi+2m_0)}\Phi^{\rm{LIS}}\left(E+\frac{|Z|e}{A}\phi\right),
\end{equation}
where $m_0$ represents the mass of a CR particle,
$\Phi=\frac{dN}{dE}$ represents the differential intensity of CR, and $\Phi^{\rm{LIS}}$ represents the LIS outside the heliosphere.

According to Eq.~(\ref{eq:FFA}), the modulated CR spectrum at a specific time depends on both the LIS and the temporal modulation potential $\phi(t)$.
However, the degeneracy between them is difficult to break~\citep{Maurin:2014bva}.
Direct measurements of the LIS are inherently limited, as most space-based experiments operate in the vicinity of Earth. Notable exceptions are Voyager 1 and 2~\citep{Cummings:2016pdr, 2019NatAs...3.1013S}, which have measured the CR spectrum outside the Solar cavity at energies $\lesssim0.1$~GeV.
CR spectral results were collected over different periods, and it is common practice to adopt specific values of $\phi$ that best fit the data averaged over given periods for different experiments. However, if the spectral shape changes rapidly with time, the averaged spectrum should not be directly modulated from the LIS, otherwise the result would be biased.

Recently, the AMS-02 collaboration has provided precise measurements of $p$, $\bar{p}$, $e^-$, and $e^+$ per Bartels rotation (BR, a 27-day periods of rotations of the sun)~\citep{AMS:2025npj}, offering valuable insights into the temporal variations of solar modulation.
They found that the variation magnitude of the $\bar p$ temporal flux is significantly smaller than those of $p$, $e^-$, and $e^+$ at low energies, which is related to its larger spectral index $\gamma$. This relation can be well understood within the framework of the FFA model.

\begin{figure}[ht!]
\includegraphics[width=0.45\textwidth,trim=0 0 0 0,clip]{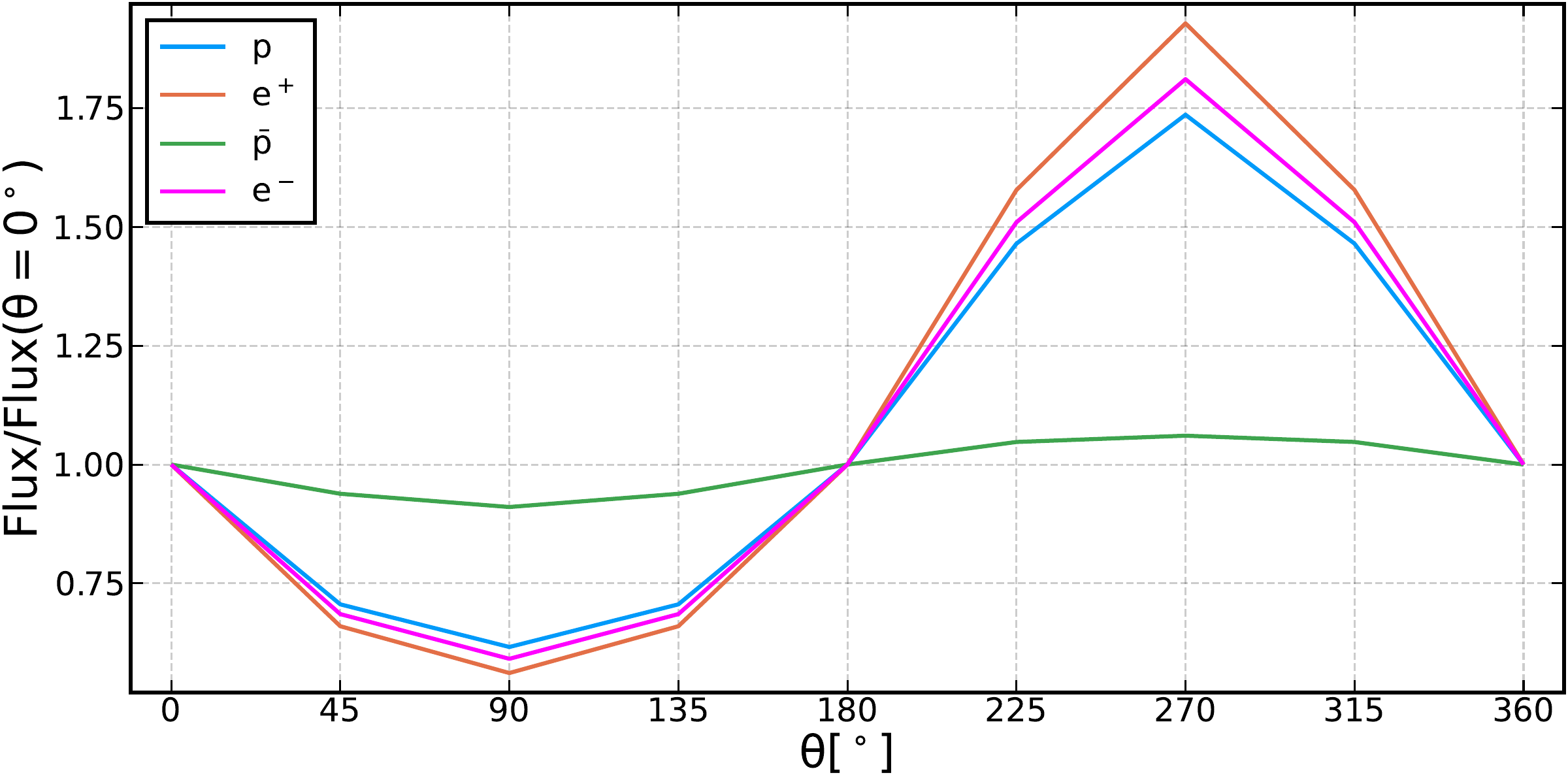}
\caption{The cyclical evolutions of modulated fluxes (divided by the fluxes where $\theta=0^\circ$) at 1.2 GV compared across different CR species. The modulation potential is assumed to change with the phase angle, $\phi(\theta)=(0.2\sin(\theta)+0.6)$~GV.
\label{fig:thetaphi}}
\end{figure}

As a qualitative demonstration, we calculate the LIS of $p$, $\bar p$, $e^-$, and $e^+$ using the numerical GALPROP framework~\citep{Strong:1998pw, Strong:1998fr, Porter:2021tlr}. The spectral indices of them at $\sim1.2$~GV exhibit significant differences: $\gamma_{e^+}(-3.3)< \gamma_{e^-}(-2.75) <\gamma_p(-1.5) < \gamma_{\bar p}(3.0)$.
Since the temporal evolutions of the observed fluxes~\citep{AMS:2025npj} follow an approximately cyclical pattern over an 11-year Solar cycle, we assume that the modulation potential can be described by $\phi(\theta)=(0.2\sin(\theta)+0.6)$~GV.
The modulated fluxes of these CRs at 1.2~GV are shown in Figure.~\ref{fig:thetaphi}.
The variation magnitude of $\bar p$ flux is significantly smaller than those of $p$, $e^-$, and $e^+$, even though they are modulated with the same $\phi$. This behavior can be understood through Eq.~(\ref{eq:FFA}). If $\gamma$ is large enough and $\Phi^{\rm{LIS}}\left(E+\frac{|Z|e}{A}\phi\right)>\Phi^{\rm{LIS}}\left(E\right)$, the flux of modulated spectrum would not be reduced significantly.
Moreover, FFA successfully predicts the relation between spectral index $\gamma$ and the variation magnitude $M$. Specifically, the condition $\gamma_{e^+}<\gamma_p$ results in $M_{e^+}>M_p$. 


\section{the Non-LIS method} \label{sec:Non-LIS}

The regularity of solar modulation can be effectively separated by using the Non-LIS method of the FFA, as introduced in~\citet{Corti:2019jlt} and~\citet{Long:2024nty}. By defining the relative modulation potential $\Delta\phi=\phi(t)-\phi(t_0)$, where $t_0$ is an arbitrary reference time, the modulated flux at time $t$ can be derived as:
\begin{equation}\label{eq:nonLIS}
\begin{aligned}
     \Phi\left(E,t\right)=&\frac{E(E+2m_0)}{(E+\frac{|Z|e}{A}\Delta\phi)(E+\frac{|Z|e}{A}\Delta\phi+2m_0)}\\
     &\times\Phi\left(E+\frac{|Z|e}{A}\Delta\phi,t_0\right)\,.
\end{aligned}
\end{equation}
This method demonstrates that the CR spectrum at any given time can be transformed into the spectrum at another time through modulation by $\Delta\phi$, without requiring knowledge of the LIS. 

If Eq.~(\ref{eq:FFA}) is simplified as $\Phi\left(\phi\right)=FFA(\Phi^{\rm{LIS}},\phi)$, Eq.~(\ref{eq:nonLIS}) can be rewritten as:
\begin{equation}
\begin{aligned}
 \Phi\left(\phi_0+\Delta\phi\right)=&FFA(\Phi\left(\phi_0\right),\Delta\phi)\\
 =&FFA(FFA(\Phi^{\rm{LIS}},\phi_0),\Delta\phi)\,.
 \end{aligned}
\end{equation}
The reference modulation potential $\phi(t_0)=\phi_0$ only represents the difference between $\Phi^{\rm{LIS}}$ and $\Phi\left(\phi_0\right)$, which is independent of the modulation effect from $t_0$ to $t$.
We can also demodulate any spectrum modulated by a potential $\phi$ to its LIS by applying a modulation of $-\phi$:
\begin{equation}\label{eq:demodulate}
 \Phi^{\rm{LIS}}=FFA(\Phi\left(\phi\right),-\phi)\,.
\end{equation}

As suggested in \citet{Wang:2019xtu, Wang:2020iev}, the traditional FFA model is reliable for describing solar modulation effects during periods of minimal solar activity. We select $t_0$ in these periods, during which the modulation potential is at its smallest throughout the solar cycle and is energy-independent. The reference energy spectrum at $t_0$ is modulated by $\Delta\phi(t)$ to fit the spectra in each period. Therefore, based on the time-varying measurements of CRs, we can obtain the $\Delta\phi(t)$ for different particle species.

\section{Results} \label{sec:result}
\subsection{the temporal variation of modulation potential}\label{sec:result1}

In this section, we employ the Non-LIS method of the FFA to calculate the relative modulation potential $\Delta\phi$ for each time period, utilizing the latest AMS-02 data for $p$, $\bar p$, $e^-$, and $e^+$~\citep{AMS:2025npj}. 
The dataset spans an 11-year period, encompassing BRs from 2426 to 2575 (May 2011 to June 2022).

The AMS-02 measurements indicate that the solar minimum period, during which CRs experience minimal modulation effects, is consistent across different particles. This can be seen from that the spectra of $p$, $\bar p$, $e^-$, and $e^+$ all reach their highest flux between Mar 2020 and May 2020 (BR2545-2547). This period is expected to coincide with the lowest solar activity within the solar cycle.  
To reduce the nuisances of the data, we calculate the average spectrum across BR2545, BR2546, and BR2547, which serves as the reference spectrum $\Phi_0$. Note that the solar minimum period selected here BR2545–2547 differs from that in \citet{Long:2024nty} BR2502-2505.
It seems that the period BR2502-2505 coincides with the onset of a sharp activity bump around BR2510 during the LSA period. This bump may be associated with short-term solar activities, rendering it less representative as a reference spectrum.

The reference spectrum is subsequently modulated to align with the spectra observed in other BRs.
For the modulation process, we utilize the cubic spline method to interpolate the fluxes within the rigidity range of $\sim1-42$~GV\footnote{There are only 11 energy bins in this range, hence the simple power-law interpolation may lead to an underestimation or overestimation of the fluxes between the measured data points.}. A power-law extrapolation is applied to predict the fluxes outside this rigidity range, a reliable approach given that the spectral index above 30~GV is almost energy-independent. Note that when transforming the  observed results to the fluxes at the central rigidity in each rigidity bin, a slight modification should be applied. For further details, see Appendix.~\ref{sec:bin_modify}.

\begin{figure}[ht!]
\includegraphics[width=0.45\textwidth,trim=0 0 0 0,clip]{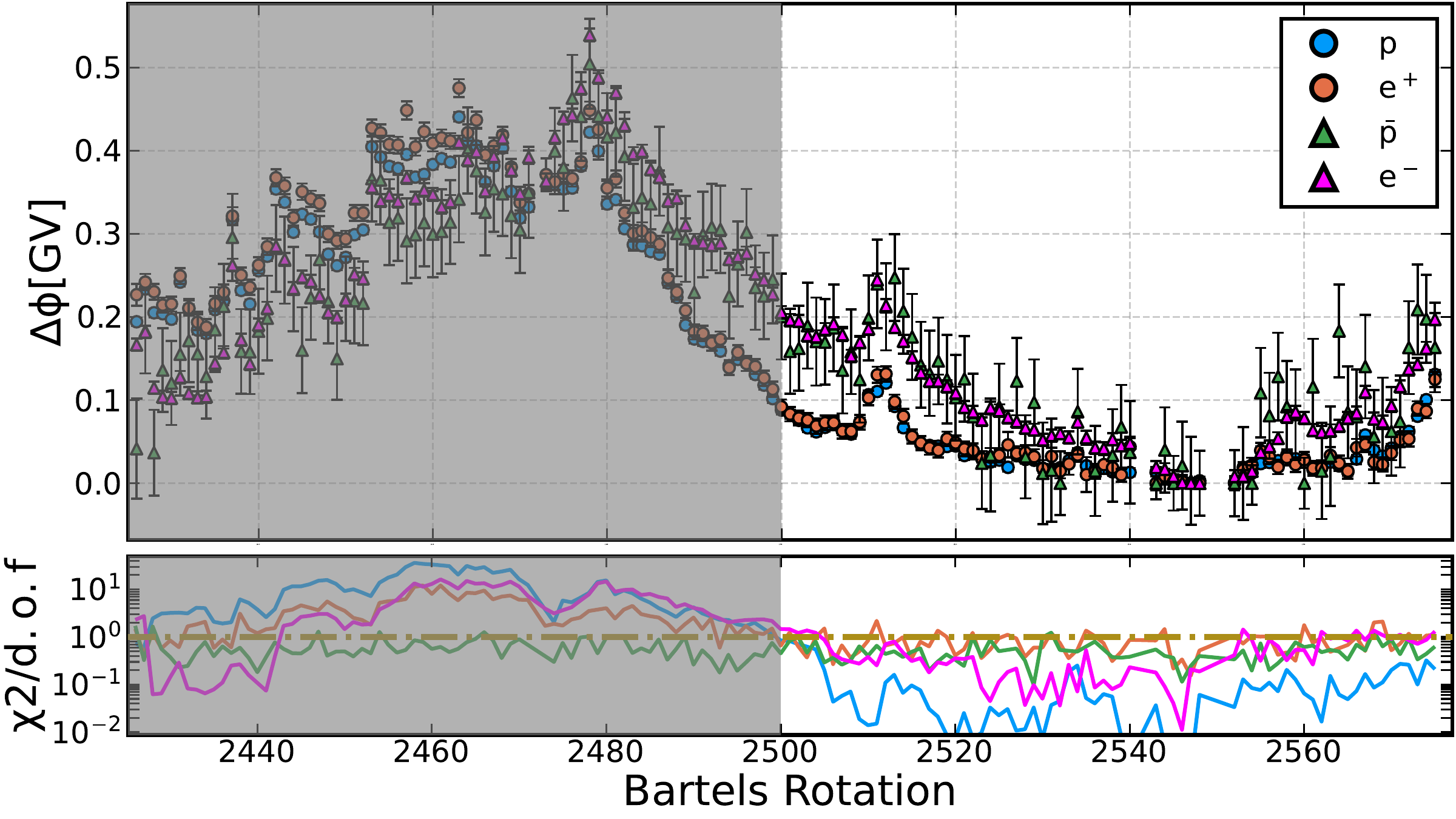}
\caption{The relative modulation potential $\Delta\phi$ and corresponding $\chi^2$/d.o.f in each BR for $p$, $\bar{p}$, $e^-$, and $e^+$. The shaded area denotes the HSA period, where the $\chi^2$/d.o.f of $p$ is larger than 1.
\label{fig:dphichisq-bartel}}
\end{figure}

A least-square analysis is performed by fitting the observations with the modulated spectrum $FFA(\Phi_0,\Delta\phi)$.
In Figure.~\ref{fig:dphichisq-bartel}, we illustrate the results for $p$ (blue), $\bar p$ (green), $e^-$ (magenta), and $e^+$ (orange) as a function of BR period.
In the top panel, the $\Delta\phi$ for $p$ and $e^+$ exhibit consistent temporal variations, as do those for $\bar p$ and $e^-$. This indicates that the features of $\Delta\phi$ (and consequently $\phi$) over the solar cycle are charge-sign dependent. Furthermore, the solar modulation of particles with the same charge sign during the LSA period can be described by a unified FFA model.
In the bottom panel, the FFA model well describes the solar modulation during the LSA period between BR2500 and BR2575, as evidenced by the $\chi^2$/d.o.f value for $p$ (blue line) being significantly less than 1.
However, the high solar activity (HSA) period between BR2426 and BR2499 proves challenging to fit, implying that the traditional FFA model over-simplifies some physical processes during these periods. Some modified FFA models could be used to reproduce the spectra during the HSA period better~\citep{Zhu:2025qqx,Cholis:2020tpi, Kuhlen:2019hqb}.
Despite this limitation, the validity of the FFA model during the LSA period is sufficient for estimating the LIS of $p$ and $e^+$ in the next subsection.

A global time shift in $\Delta\phi$ between particles of opposite charge signs can be seen between BR2426 and BR2490, with the effects on negatively charged particles appearing delayed. This feature could be used to study the hysteresis found by AMS-02 and the embedded magnetic field within the heliosphere~\citep{Lipari:2023sja}.
We also notice that the $\Delta\phi$ values for $e^+$ are systematically larger ($\sim3\sigma$) than those of $p$ from BR2440 to BR2485. Similarly, a difference is found between $\bar p$ and $e^-$ from BR2450 to BR2490, with the latter systematically larger.
This implies that the solar modulation effect during the HSA period is stronger for leptons. However, given the worse $\chi^2$/d.o.f during this period, the FFA model may fail during the polarity reversal of the magnetic field at the solar maximum.

As shown in Figure.~\ref{fig:dphichisq-bartel}, the $\Delta\phi$ uncertainty of $\bar p$ is overwhelmingly larger than those of other particles. This is caused by two reasons: 1) The data uncertainties of $\bar p$ ($\sim \mathcal{O}(10\%)$) are larger than those of $p$ and $e^-$, also resulting in relatively smaller reduced $\chi^2$ for $\bar p$. 2) The spectrum of $\bar p$ is less sensitive to changes in the modulation potential (see Figure.~\ref{fig:thetaphi}), and small nuisances of data can significantly impact the fitting result for $\Delta\phi$.

Other discussions are presented as follows. In Appendix.~\ref{sec:R-dphi}, we individually fit the $p$ data at each energy bin to analyze the rigidity dependency of $\Delta\phi$. During the HSA period, $\Delta\phi$ becomes larger at lower energies, leading to poor reduced $\chi^2$ in Figure.~\ref{fig:dphichisq-bartel}.
As one of the FFA assumptions, the diffusion coefficient is proportional to rigidity $\kappa\propto R$.
Hence, the energy dependency of $\kappa$ might be changed by strong solar activities during the HSA period \citep{Wang:2020iev}, resulting in a rigidity-dependent potential.

In Appendix.~\ref{sec:other_dphi}, we use AMS-02 data~\citep{AMS:2025pgu} to calculate $\Delta\phi$ for more positively charged particles, ranging from He to O. The temporal variation features of them are consistent, especially during the LSA period, implying that an identical modulation potential can be assumed for all positively charged particles. 


\subsection{the LIS of proton and positron}\label{sec:result2}
Voyager 1~\citep{Cummings:2016pdr} has measured the $p$ spectrum outside the heliosphere from 3~MeV to 300~MeV, which is useful for determining the LIS and can serve as the reference spectrum at low energies. Combined with the whole reference spectrum $\Phi_0$ during the solar minimum period above 1~GV (roughly 0.75~GeV for LIS), we can derive the reference modulation potential $\phi_0$ and the $p$ LIS.

\begin{table}
\centering
\caption{The best-fit injection parameters and modulation potential for $p$ during the solar minimum period ($\chi^2$/d.o.f=8.55/18). \label{table: injection}}
\begin{tabular}{cccccccc}
\hline
\hline
$\nu_0$ & $R_{br0}$ & $\nu_1$ & $R_{br1}$ & $\nu_2$ & $R_{br2}$ & $\nu_3$ & $\phi_0$\\
 & (GV) & & (GV) & & (GV) & & (GV)\\
\hline
1.593 & 0.3 & 2.070 & 8.4 & 2.283 & 22.2 & 2.368 & 0.32 \\
\hline
\end{tabular}
\end{table}

To generate the $p$ LIS for fitting,
we employ the GALPROP framework, as detailed in our previous work~\citep{Zhao:2024qbj, Zhao:2025szm}.
The propagation parameters are fixed to values determined by fitting the spectra of B, C, O, $\rm^{7}Be$, and $\rm^{10}Be$.
We assume that the injection spectrum of protons follows a broken power law, with three break rigidities at around 0.5~GV ($R_{br0}$), 8~GV ($R_{br1}$), and 20~GV ($R_{br2}$) to describe the spectral structure measured by AMS-02 and Voyager 1. $\phi_0$ is defined as the best-fit value required to modulate the $p$ LIS to the reference spectrum $\Phi_0$.

\begin{figure}[ht!]
\includegraphics[width=0.45\textwidth,trim=0 0 0 0,clip]{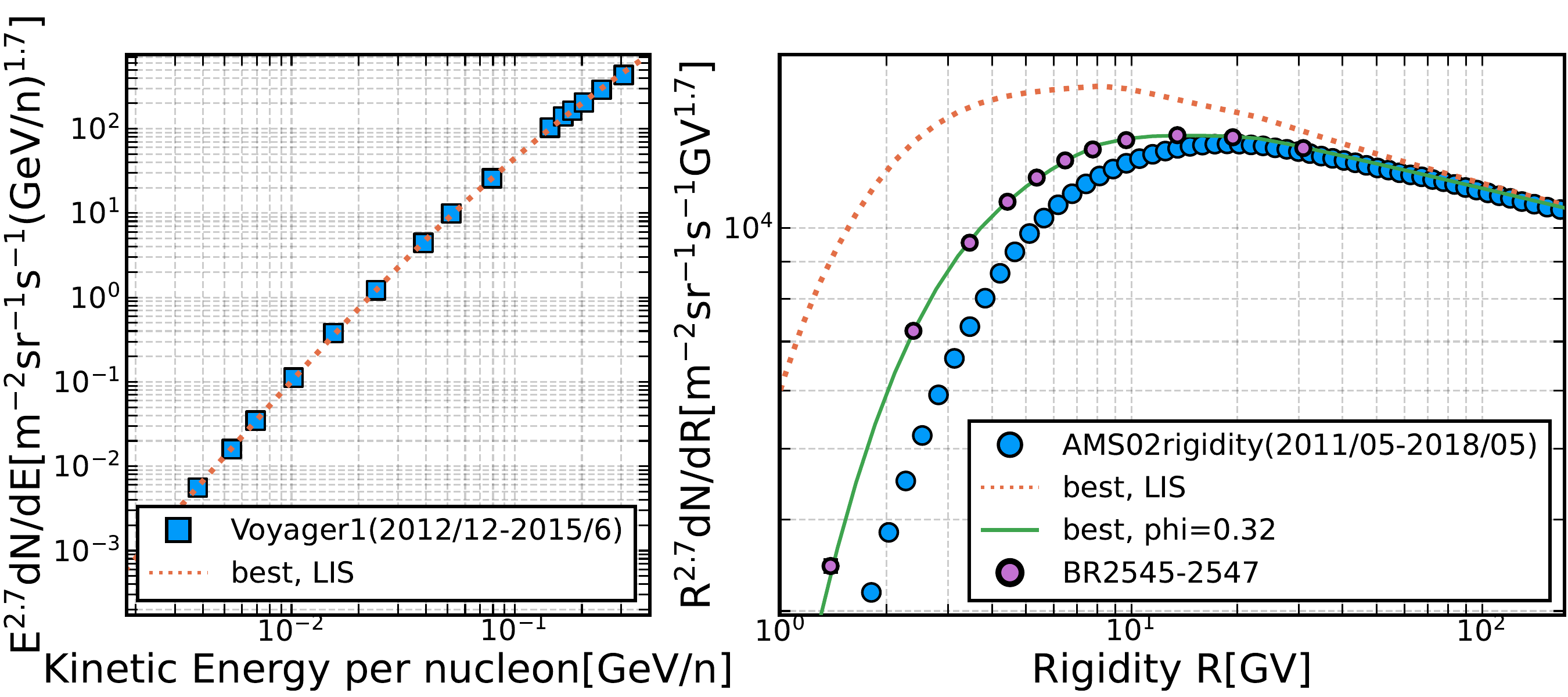}
\caption{The best-fit spectrum of $p$ and corresponding LIS. Also shown are measurements from Voyager 1~\citep{Cummings:2016pdr}, AMS-02~\citep{Aguilar:2021tos}, and the average spectrum~\citep{AMS:2025npj} during BR2545–2547.
\label{fig:p-LIS}}
\end{figure}

By fitting the reference spectrum $\Phi_0$, we obtained the best-fit values of the injection parameters as well as $\phi_0$.
These results are listed in Table~\ref{table: injection} and the corresponding best-fit $p$ LIS is shown in Figure.~\ref{fig:p-LIS}. Our result $\phi_0=$ 0.32~GV is smaller than those from the Neutron Monitor Data Base\footnote{http://www.nmdb.eu.} 0.4~GV$<\phi$<0.5~GV during the same period. Note that the neutron monitor results given by~\cite{Ghelfi:2016pcv} are $\sim0.1$~GV larger than those of~\cite{2005JGRA..11012108U}, with the latter expected to align more closely with our results.

The results presented in Sec.~\ref{sec:result1} indicate that the solar modulation of CR particles with the same charge sign during the LSA period can be effectively described by a unified FFA model.
Therefore, we adopt the same reference modulation potential $\phi_0=0.32$~GV for $e^+$. With the temporal variation $\Delta\phi(t)$ obtained in Sec.~\ref{sec:result1}, the total modulation potential $\Delta\phi(t)+\phi_0$ in each BR is determined.
According to Eq.~(\ref{eq:demodulate}), the $e^+$ spectra from AMS-02~\citep{AMS:2025npj} can be demodulated by applying $-(\Delta\phi(t)+\phi_0)$ to acquire the LIS.
To avoid incorrect assumptions from extrapolation and interpolation, especially at low energies where the spectral index $\gamma$ changes rapidly, we demodulate the data for each energy bin one by one. For example, the central energy of the first bin, 1.386~GV, is demodulated to (1.386+$\Delta\phi(t)+\phi_0$)~GV.

\begin{figure}[ht!]
\includegraphics[width=0.45\textwidth,trim=0 0 0 0,clip]{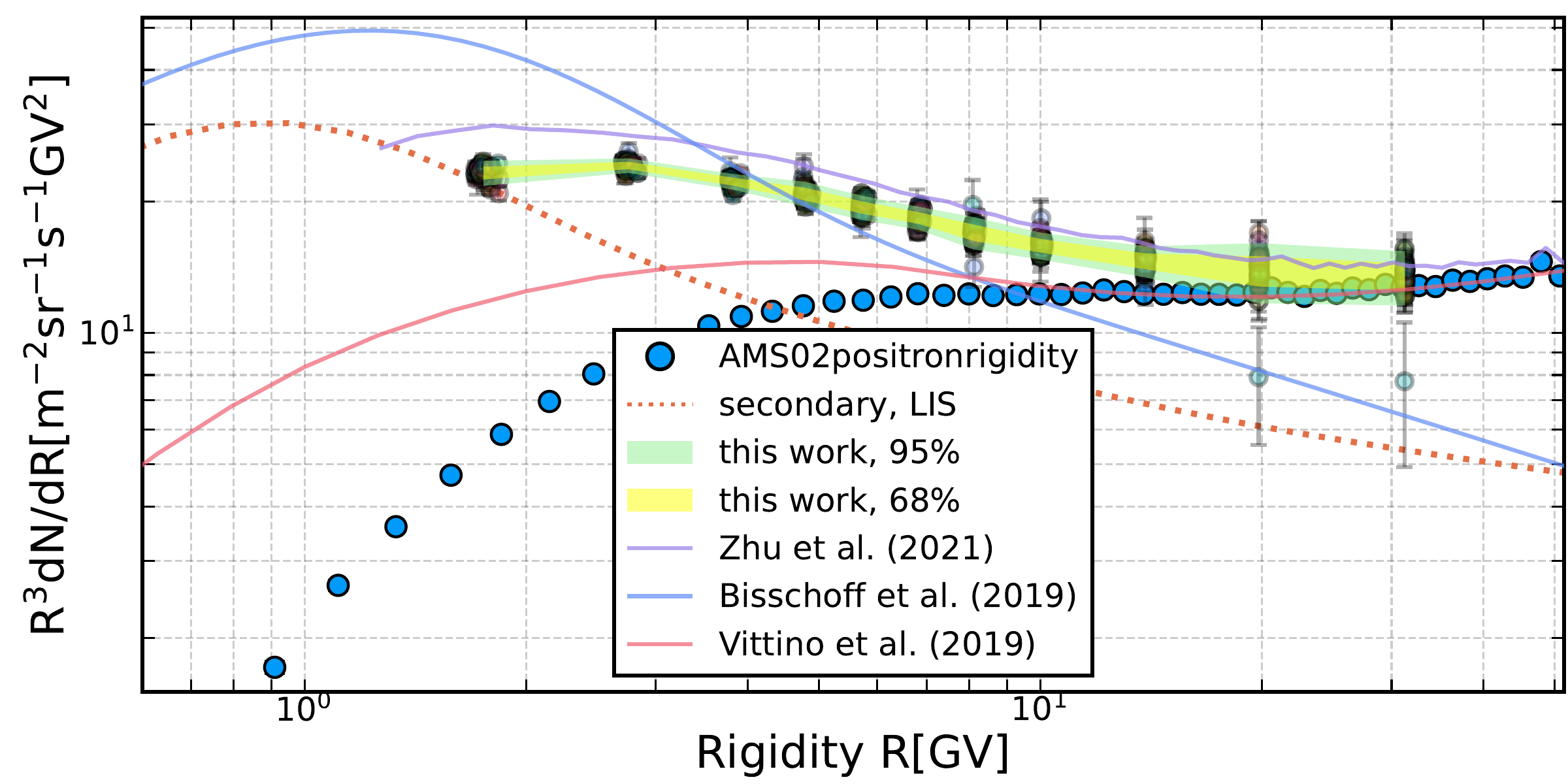}
\caption{Comparison of our derived LIS of $e^+$ (95\% and 68\% confidence interval) with previous works~\citep{Zhu:2020koq,Bisschoff:2019lne,Vittino:2019yme}. Also shown are measurements from AMS-02~\citep{Aguilar:2021tos} and our predicted secondary $e^+$ spectrum (red dotted line) calculated with the GALPROP framework.
\label{fig:e-LIS}}
\end{figure}

As a result, 11 groups of demodulated fluxes (represented as dots) are shown in Figure.~\ref{fig:e-LIS}, which directly reflect the constraints imposed by the AMS-02 data from 11 energy bins.
The 68\% and 95\% confidence intervals (yellow and green band) are derived from the demodulated fluxes of $70+$ periods from BR2500 to BR2575, providing a reliable estimation of the LIS. The 68\% confidence interval of the $e^+$ LIS is listed in Table.~\ref{tab:e+LIS} as a reference.
The LIS of previous works with different methods and assumptions~\citep{Zhu:2020koq,Bisschoff:2019lne,Vittino:2019yme} shows relatively large differences from ours. Surprisingly, our result, derived by assuming that $p$ and $e^+$ have the same modulation potential, is close to that of~\citet{Zhu:2020koq} (purple line), which is based on a non-parametric description of the LIS and fitted to the long-term average $e^+$ data.
Additionally, we present a theoretical expectation for secondary $e^+$ production calculated with GALPROP (red dotted line).
The propagation and injection parameters are the same as those we used to generate the $p$ LIS mentioned above. 
The numerical solution of GALPROP takes account of the dominant losses, such as the synchrotron losses on the Galactic magnetic field and the inverse Compton losses on the interstellar
radiation fields. Additionally, adiabatic, bremsstrahlung, and ionization losses, which impact the prediction around a few GeV, are also taken into consideration. The interstellar radiation field model utilized in this study is the default GALPROP one~\citep{Porter:2005qx}. Synchrotron
energy losses are computed based on a regular magnetic field~\citep{Pshirkov:2011um}, along with a random component modeled according to~\citep{Sun:2007mx}.
We employed the new $e^{\pm}$ production cross section model developed by~\citep{Orusa:2022pvp} based on the latest collider data.
The theoretical expectation for secondary $e^+$ is significantly lower than our LIS obtained through de-modulation.
This discrepancy implies that, beyond the production of positrons from the decay of charged pions and kaons generated in collisions of CRs with interstellar gas, additional $e^+$ sources are required above 3~GV. Potential contributors include pulsar wind nebulae, dark matter annihilation or decay, etc.~\citep{Hooper:2008kg, Arkani-Hamed:2008hhe, Yin:2008bs, Yuan:2013eja, Lin:2014vja}.

\begin{table}
\caption{\label{tab:e+LIS}The 68\% confidence interval of the $e^+$ LIS.}
\begin{ruledtabular}
\begin{tabular}{c|c}
 Rigidity [GV]& $dN/dR$ [$\mathrm{m}^{-2}\mathrm{sr}^{-1}\mathrm{s}^{-1}\mathrm{GV}^{-1}$]\\ \hline
  1.7486  & [4.19943,    4.48942]\\
  2.75093 & [1.13825,      1.18325]\\
  3.8183  & [0.390352,     0.407043]\\
  4.79213 & [0.182497,     0.19522]\\
  5.72878 & [0.0996197,    0.106246]\\
  6.83065 & [0.0557088,    0.0592442]\\
  8.11687 & [0.0303428,    0.0330778]\\
 10.0211  & [0.0151604,    0.0162882]\\
 13.8759  & [0.00523331,   0.00568663]\\
 19.8175  & [0.00163887,   0.00192276]\\
 31.2712  & [0.000408342,  0.000471875]\\
\end{tabular}
\end{ruledtabular}
\end{table}

Determining the LIS of antiprotons and electrons is challenging for two main reasons.
Firstly, their temporal variations are  inconsistent with those of protons, implying that their reference modulation potential cannot be equal to that of protons.
Secondly, there are no available LIS measurements for $\bar p$ or $e^-$, hence the degeneracy between injection parameters and reference modulation potential $\phi_0$ cannot be broken.
Voyager 1~\citep{Cummings:2016pdr} has measured the total lepton spectrum $\Phi_{e^+}+\Phi_{e^-}$ from 3~MeV to 30~MeV, and the LIS contribution of $e^+$ can be subtracted.
However, the energy loss of electrons is dominant over their escape from the Galaxy, making the spectral index changes rapidly with energy. Hence, it is difficult to extrapolate Voyager's data to estimate the LIS at higher energies. We leave further discussions in this regard to future studies.

\section{Conclusion} \label{sec:Conclusion}

The widely used FFA model offers a simplified yet effective  framework for describing variations in CR spectra within the heliosphere through the modulation potential, $\phi$. 
In this study, we utilize the Non-LIS method of FFA to individually calculate the temporal variation of the relative modulation potential $\Delta\phi(t)$, taking the spectrum during the period of minimal solar modulation as a reference. Notably, this methodology does not rely on prior knowledge of the LIS.

Our analysis demonstrates that the CR spectra of $p$, $e^+$, $e^-$, and $\bar p$ measured by AMS-02 can be effectively characterized by the FFA model during the LSA period. Notably, the derived $\Delta\phi(t)$ for both $p$ and $e^+$ (and similarly for $e^-$ and $\bar p$) exhibits perfect consistency. We conclude that the solar modulation of particles with the same charge sign during the LSA period can be described by a unified FFA model. 

On the other hand, during the HSA period, the basic FFA model fails to adequately fit the data, suggesting the need for a rigidity-dependent $\Delta\phi(t)$ that increases at lower energies. This deviation may arise from the rigidity dependence of the diffusion coefficient in the heliosphere during the HSA period, departing from the assumed $\kappa\propto R$. Despite this limitation, the results during the HSA period do not affect the determination of the LIS.

Based on these findings, it is reasonable to assume that the modulation potential of $e^+$ aligns with that of $p$ during the LSA period. We determine the LIS of $p$ and the reference modulation potential $\phi_0$ for positively charged particles by fitting the reference spectrum, based on data from Voyager 1 and AMS-02. Using this $\phi_0$ and the relative modulation potential $\Delta\phi(t)$, we derive the total potential $\phi_0+\Delta\phi(t)$ for each BR period. The AMS-02 spectra of $e^+$ during the LSA period are demodulated to derive their common LIS, without necessitating assumptions about their origin and propagation mechanisms. Notably, above $3$~GV, the derived $e^+$ LIS significantly exceeds the predictions of the traditional CR model, which only considers the secondary positrons generated in collisions of CRs with interstellar gas. This suggests that primary $e^+$ contributions should be assumed. 

In addition, we find that the temporal variations of $\Delta\phi(t)$ for other positively charged heavy nuclei are consistent with those of $p$ and $e^+$ during the LSA period, further supporting our conclusions. The LIS of these heavy CRs can be determined using the same method applied in this study.

The limitation of the FFA has been studied by~\citet{2004JGRA..109.1101C}. The modulation potential in the FFA model is regarded as a proxy for the time variations of the CR modulation rather than a physical quantity~\citep{Tomassetti:2022fal}.
The large discrepancy of the FFA potential between positively and negatively charged CRs might be related to the more intricate mechanisms, such as the drift effect~\citep{Cholis:2015gna,Aslam:2023gjv}.
Hence, it is meaningful to analyze the AMS-02 data with more realistic models to verify our results.

\begin{acknowledgments}
   This work is supported by the National Key R\&D Program Grants of China under Grant No. 2022YFA1604802, 
the National Natural Science Foundation of China under Grants No. 12175248, No. 12342502, No. 12105292, and No. 12393853.  
\end{acknowledgments}

\appendix
\section{Rigidity dependence of $\Delta\phi$} \label{sec:R-dphi}

\begin{figure}[ht!]
\centering
\includegraphics[width=0.7\textwidth,trim=0 0 0 0,clip]{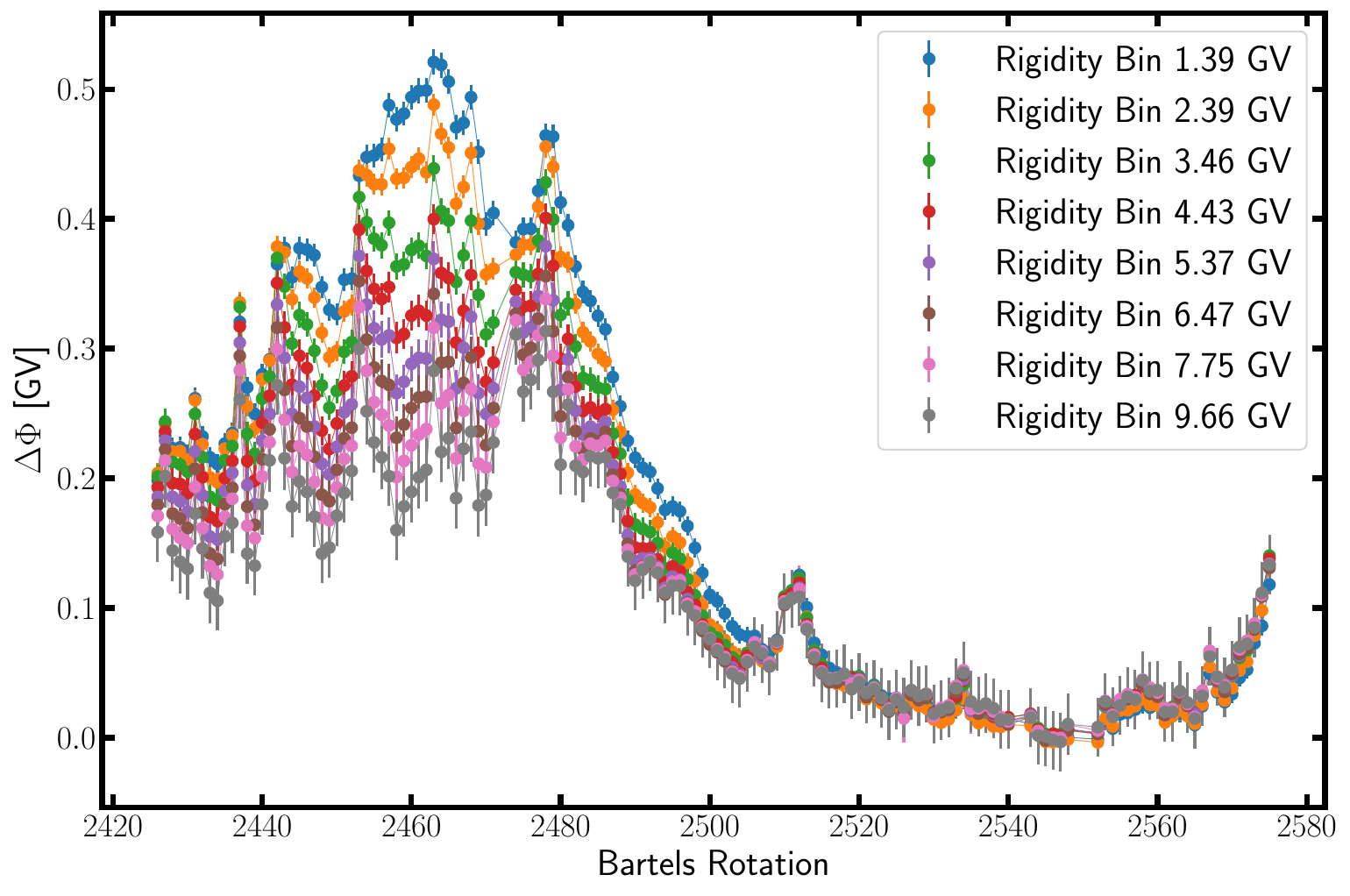}
\caption{The relative modulation potentials for each energy bin, based on the $p$ data of AMS-02~\citep{AMS:2025npj}.
\label{fig:delta_phi_rigidity}}
\end{figure}

To assess the validity of the FFA in light of the AMS-02 data~\citep{AMS:2025npj}, we employ the Non-LIS method to determine the relative solar modulation potential $\Delta\phi$ of $p$ individually for each energy bin. 

The fitted results are presented in Figure.~\ref{fig:delta_phi_rigidity}. 
As the large uncertainties in high-energy bins preclude meaningful constraints, we illustrate the results of the first eight bins.
During the HSA period, $\Delta\phi$ exhibits distinct values across rigidity bins with a clear trend: lower rigidity bins correspond to larger $\Delta\phi$, indicating stronger modulation. Conversely, during the LSA period, $\Delta\phi$ remains nearly uniform across rigidity bins, signifying the absence of rigidity dependence. 
Since the basic FFA fails during the HSA period, the fitted $\Delta\phi$ values in Figure.~\ref{fig:dphichisq-bartel} might not be interpreted as meaningful results and should instead be analyzed using more sophisticated models. 

In addition, the $p$ LIS derived in Sec.~\ref{sec:result1} does not perfectly fit the data in the lowest energy bin. The measured fluxes at $1.39$~GV are systematically larger ($>2\sigma$) than the calculated spectra (modulated by $\phi(t)$) during the same period, and this discrepancy only vanishes around the BR2510, where the sharp bump structure in $\Delta \phi$ lies.
As suggested in \citet{2004JGRA..109.1101C}, even during the LSA period, the FFA model may be overly simplistic for properly modulating the spectrum at low energies, compared with the full numerical solution. Corti's work~\citep{Corti:2015bqi} found a bump structure around 1 GV by fitting the PAMELA data, which may also be due to the limitation of the FFA at low energies.

\section{Results of $\Delta\phi$ for other CR particles}\label{sec:other_dphi}
To assess the validity of the consistent $\Delta\phi$ results for positively charged particles obtained in Figure.~\ref{fig:dphichisq-bartel}, we calculate the relative modulation potential $\Delta\phi$ for several heavy nuclei, including He, Li, Be, B, C, N, and O, using another dataset provided by AMS-02~\citep{AMS:2025pgu}.

\begin{figure}[ht!]
\centering
\includegraphics[width=0.7\textwidth,trim=0 0 0 0,clip]{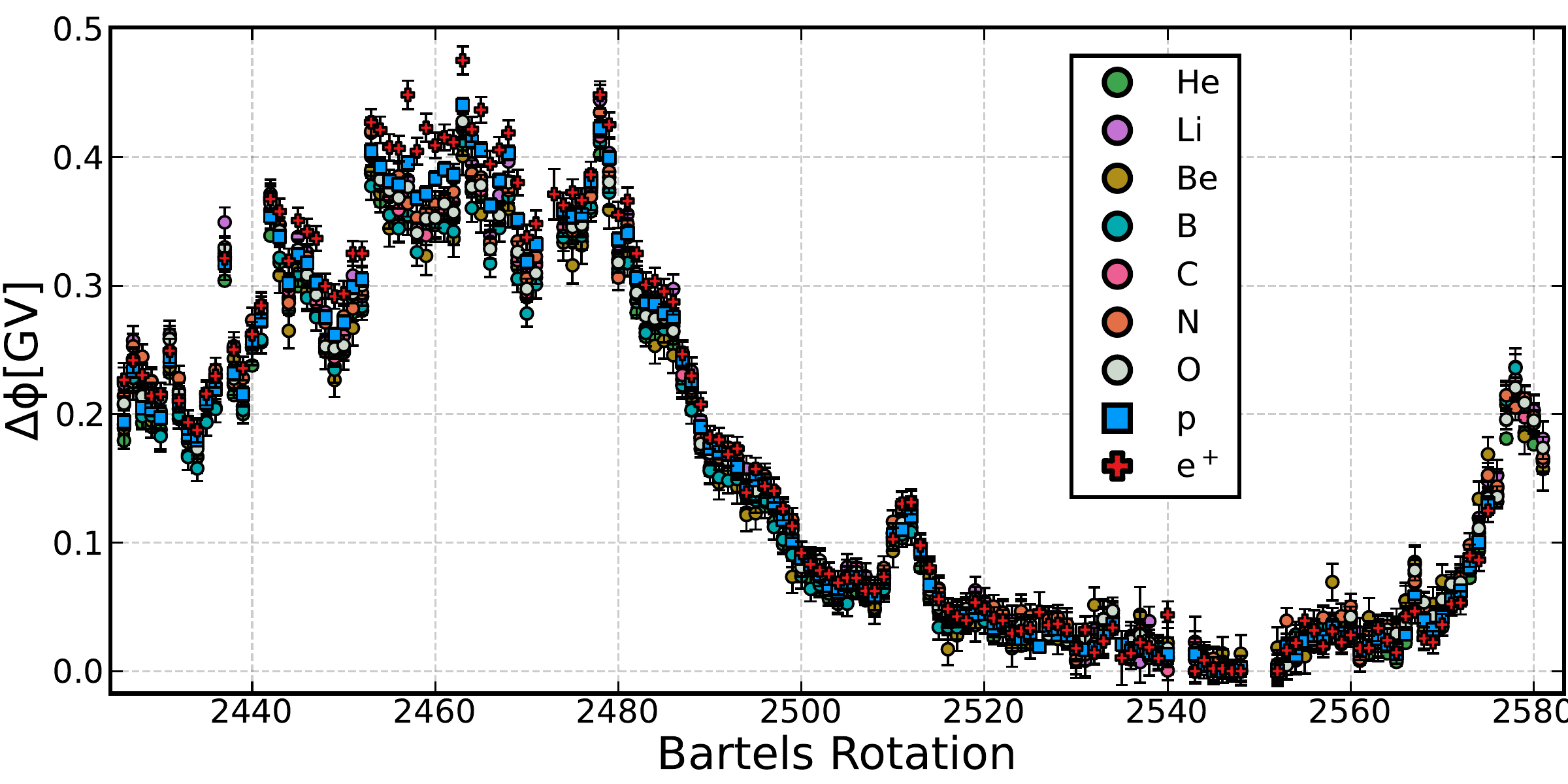}
\caption{The relative modulation potentials for positively charged CRs, based on the AMS-02 data~\citep{AMS:2025pgu}. 
\label{fig:dphiALL}}
\end{figure}

In Figure.~\ref{fig:dphiALL}, the temporal variations of $\Delta\phi$ are presented for all positively charged CRs.
The solar minimum periods for these nuclei are identified between BR2545 and BR2548, closely aligning with that of $p$.
During the LSA period, the temporal variations of $\Delta\phi$ for these particles are nearly consistent, whereas during the HSA period, $\Delta\phi$ of $e^+$ is globally larger than that of other nuclei. $\Delta\phi$ of $p$ is also slightly larger than those of heavy nuclei  around BR2465.

\begin{figure}[ht!]
\includegraphics[width=0.5\textwidth,trim=0 0 0 0,clip]{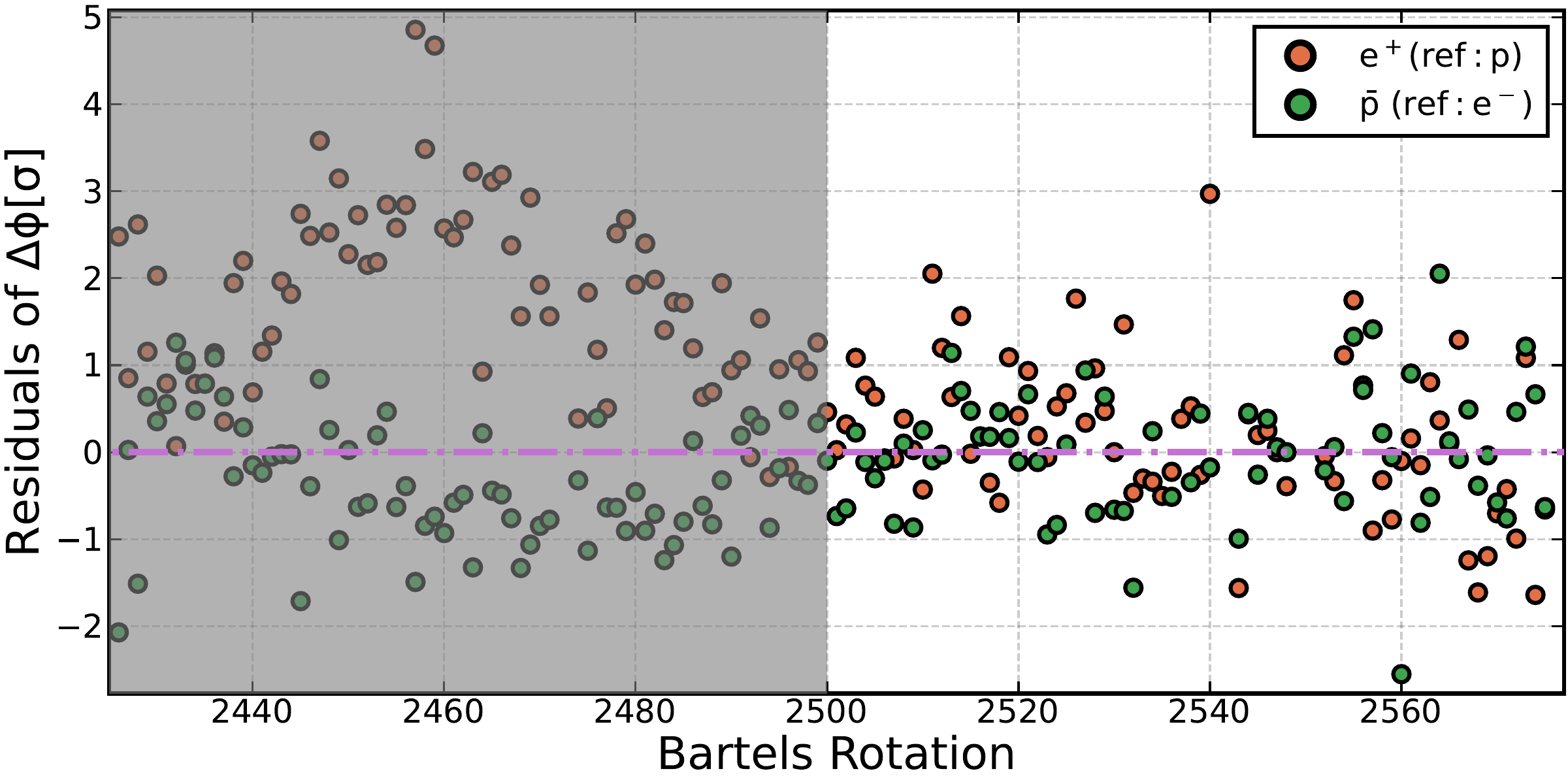}
\includegraphics[width=0.5\textwidth,trim=0 0 0 0,clip]{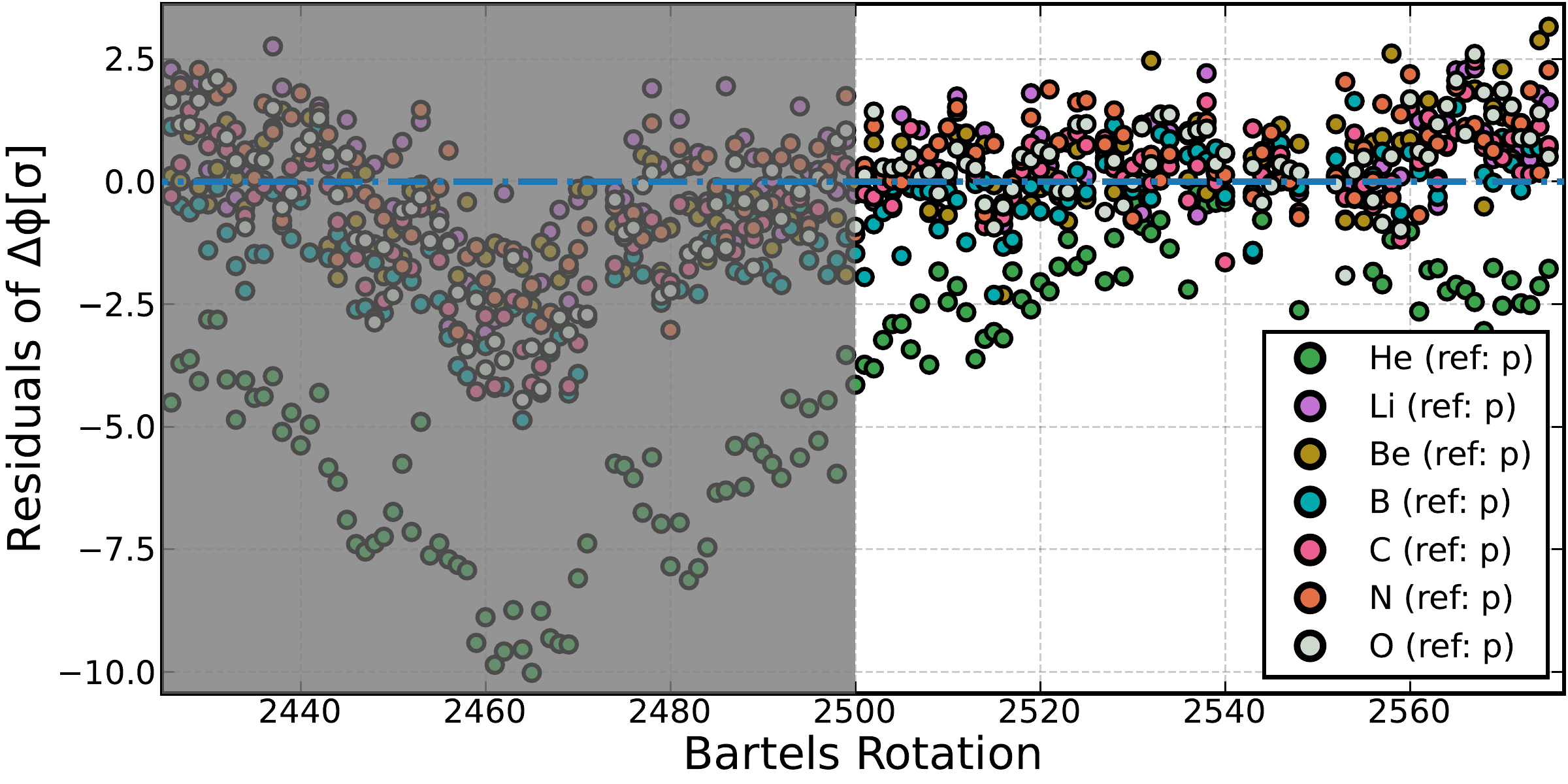}
\caption{The residuals of relative modulation potentials in each BR. Left: Comparison for particles with the same charge sign. Right: Comparison for positively charged CR nuclei.
\label{fig:res-bartel}}
\end{figure}

In Figure.~\ref{fig:res-bartel}, we calculate the residuals of $\Delta\phi$ as a function of time comparing different nuclei. The results for $p$ and $e^-$ are selected as the reference of positively and negatively charged particles, respectively, due to their small uncertainties.
In the left panel, the residuals for $e^+$ and $\bar p$ are small during the LSA period, showing no systematic bias. During the HSA period, the residuals for $e^+$ are significantly larger ($>2\sigma$), while those for $\bar p$ exhibit systematic bias around $-1\sigma$, though this is not statistically significant due to large data uncertainties.
In the right panel, the modulation potentials for all positively charged nuclei align closely during the LSA period, except for He.
Note that the residuals for He depend on the selected period of the reference spectrum, suggesting that its total modulation potential may not be systematically smaller than that of $p$.
This feature warrants further investigation using additional time-varying observations of He, such as those from PAMELA~\citep{Marcelli:2020uqv, Marcelli:2022gbn}.

A slight excess in $\Delta\phi$ is observed between BR2560 and BR2580 for nuclei from Li to O, indicating the emergence of new solar activity features at the end of the LSA period.
During the HSA period, the residuals for all heavy nuclei are significant over BR2450-2480, likely related to the common bump structure observed in Figure.~\ref{fig:dphiALL} and \ref{fig:dphichisq-bartel}. Combined with the results for $e^+$ and $\bar{p}$, this bump structure in $\Delta\phi$ might be associated with differences between leptons and nuclei, as well as the mass-to-charge ratio A/Z.

\section{Energy Bin modification}\label{sec:bin_modify}
The observations in~\citet{AMS:2025npj} are provided with 11 rigidity bins from 1~GV to 41.9~GV. The determination of the central rigidity requires careful consideration, as the width of each bin is an order of magnitude larger than that of the often-used average spectra at the same energy. 
The measured flux in the $i$th rigidity bin ($R_i, R_i+\Delta R_i$) is given by the (corrected) number of events $N_i$ divided by the bin width $\Delta R_i$, corresponding to the average integral intensity:
\begin{equation}
  \Phi_i^{data}=\frac{1}{\Delta R_i}\int_{R_i}^{R_i+\Delta R_i}\frac{dN}{dR}\, dR\,,
\end{equation}
where $\frac{dN}{dR}$ is the differential intensity of CR as a function of rigidity.
The central rigidity $\bar{R}$ can be defined as the rigidity where the differential intensity equals the average integral intensity within the rigidity bin, $\frac{dN}{dR}(\bar{R})=\Phi^{data}$. 

For convenience, the geometric mean rigidity $R'_i=\sqrt{{R_i}(R_i+\Delta R_i)}$ is often used to represent the $i$th rigidity bin.
However, the difference between $\bar{R}$ and $R'$ becomes non-negligible when the bin width is comparable to its rigidity $\Delta R_i\sim R_i$. The differential intensity at the geometric mean rigidity is given by:
\begin{equation}
  \frac{dN}{dR}(R'_i)=\frac{dN}{dR}(\bar{R_i})\times\frac{f(R'_i)\Delta R_i}{\int_{R_i}^{R_i+\Delta R_i}f(R)\, dR}\equiv\Phi^{data}_i\times k_i\,,
\end{equation}
where $f(R)=R^{\gamma}$ represents the spectral shape of the spectrum, and $k_i$ is a correction factor used to adjust the data flux to obtain the correct flux value at $R'$. To properly estimate the spectral shape $f(R)$, we use the average spectra provided in~\citet{AMS:2025npj} as references, which cover a broader energy range with significantly smaller bin widths. 
An alternative approach is to derive the central rigidity from $\bar{R}=f^{-1}(\frac{\int_{R_i}^{R_i+\Delta R_i}f(R)\, dR}{\Delta R_i})$. However, the corresponding rigidity would be difficult to use in the subsequent analysis.

The calculated correction factors $k_i$ are listed in Table~\ref{tab:bin_factor}, typically falling within 2\%, except for the highest and lowest energy bins.
Note that the calculation of $k_i$ at the lowest bin (1, 1.92)~GV relies on an extrapolation assumption of the spectral shape, where the spectral index changes rapidly with rigidity due to the solar modulation effect.
To avoid introducing significant bias, we keep the lowest rigidity bin unmodified.
The correction factors applied to the AMS-02 spectra only depend on the types of particles. Therefore, the result of $\Delta\phi(t)$ would not be changed, whether or not the correction is applied. But the fluxes of $p$ and $e^+$ LIS determined in this work rely on this modification, especially in the lowest and the highest bins with larger $k$.
As introduced in~\citet{Lafferty:1994cj}, the systematic uncertainty of this correction method relies on the knowledge of the underlying frequency distribution.
By changing the spectral shape $f(R)$ to the standard $R^{-2.7}$ power-law spectrum, and to the interpolations of measured spectra during different BRs, we calculated the uncertainty of $k_i$ for all rigidity bins, listed in Table~\ref{tab:bin_factor}.

\begin{table}
\caption{\label{tab:bin_factor}The geometric mean rigidities, the corresponding correction factors $k_i$ and their systematic uncertainties for all rigidity bins.
The values shown in brackets represent that the fluxes in the first bin are unmodified in the work.}
\begin{ruledtabular}
\begin{tabular}{c|ccccccccccc}
 $R'$ [GV]& 1.386 & 2.388 & 3.455 & 4.429&5.366&6.468&7.754&9.658&13.513&19.455&30.908\\ \hline
 $p$& 1.05[1.0] & 1.01 & 1.0 & 1.0&1.0&1.0&1.0&1.0&0.99&0.99&0.97\\
$e^+$& 1.05[1.0] & 1.01 & 1.0 & 1.0&1.0&1.0&1.0&0.99&0.98&0.99&0.96\\
$e^-$& 1.0[1.0] & 1.0 & 1.0 & 1.0&1.0&1.0&1.0&1.0&1.0&0.98&0.94\\
$\bar{p}$& 0.96[1.0] & 1.0 & 1.02 & 1.0&1.0&1.01&1.01&1.0&0.99&1.0&0.98\\
$\sigma_{syst.}$& $\pm2.24\%$ & $_{-1.46\%}^{+0.83\%}$ & $_{-0.70\%}^{+0.29\%}$ & $_{-0.15\%}^{+0.22\%}$&$_{-0.18\%}^{+0.13\%}$&$_{-0.18\%}^{+0.07\%}$&$_{-0.03\%}^{+0.13\%}$&$_{-0.10\%}^{+0.20\%}$&$_{-0.22\%}^{+0.37\%}$&$_{-0.10\%}^{+0.24\%}$&$_{-0.04\%}^{+1.16\%}$\\
\end{tabular}
\end{ruledtabular}
\end{table}

\bibliography{sample631}{}
\bibliographystyle{aasjournal}
\end{document}